\documentclass{article}
\usepackage{spconf,amsmath,graphicx}

\usepackage{enumitem}
\setlist{nosep, leftmargin=14pt}

\usepackage{mwe} 

\usepackage{multirow} 
\usepackage{makecell} 

\usepackage{xcolor} 

\usepackage{stmaryrd} 

\usepackage{graphicx}

\title{Synthetic magnetic resonance images for domain adaptation: Application to fetal brain tissue segmentation}

\name{\begin{tabular}{c}Priscille de Dumast$^{1,2}$, Hamza Kebiri$^{1,2}$, Kelly Payette$^{3,4}$,\\ Andras Jakab$^{3,4}$, Hélène Lajous$^{*,1,2}$, Meritxell Bach Cuadra$^{*,2,1}$\end{tabular}\thanks{* Hélène Lajous and Meritxell Bach Cuadra contributed equally to this work.}}

\address{$^{1}$  Department of Radiology, Lausanne University Hospital (CHUV) and \\ University of Lausanne (UNIL), Lausanne, Switzerland \\
$^{2}$ CIBM Center for Biomedical Imaging, Switzerland \\
$^{3}$ Center for MR Research, University Children’s Hospital Zurich, University of Zurich, \\ Zurich, Switzerland \\
$^{4}$ Neuroscience Center Zurich, University of Zurich, Zurich, Switzerland }

\begin{document}
%
\maketitle


\begin{abstract}
The quantitative assessment of the developing human brain \textit{in utero} is crucial to fully understand neurodevelopment. Thus, automated multi-tissue fetal brain segmentation algorithms are being developed, which in turn require annotated data to be trained. However, the available annotated fetal brain datasets are limited in number and heterogeneity, hampering domain adaptation strategies for robust segmentation. In this context, we use FaBiAN, a Fetal Brain magnetic resonance Acquisition Numerical phantom, to simulate various realistic magnetic resonance images of the fetal brain along with its class labels. We demonstrate that these multiple synthetic annotated data, generated at no cost and further reconstructed using the target super-resolution technique, can be successfully used for domain adaptation of a deep learning method that segments seven brain tissues. Overall, the accuracy of the segmentation is significantly enhanced, especially in the cortical gray matter, the white matter, the cerebellum, the deep gray matter and the brain stem.
\end{abstract}

\section{Introduction}

Morphometric measurements in the developing fetal brain are critical as many congenital disorders are related to a disrupted growth of the structures~\cite{egana-ugrinovic_differences_2013}. Magnetic resonance imaging (MRI) is a powerful tool to investigate equivocal neurological patterns. However, images are often corrupted by random movements of the fetus which cause signal drops. Super-resolution (SR) reconstruction techniques take advantage of the redundancy between orthogonal low-resolution (LR) series to reconstruct an isotropic high-resolution volume of the fetal brain with reduced intensity artifacts and motion sensitivity~\cite{gholipour_robust_2010,kuklisova-murgasova_reconstruction_2012,kainz_fast_2015,tourbier_efficient_2015,ebner_automated_2020}.
Subsequent multi-tissue segmentation is key for advanced quantitative analysis of the fetal brain~\cite{makropoulos_review_2018,khalili_automatic_2019,hong_fetal_2020,dou_deep_2020,payette_automatic_2021}. Although manual annotation is a cumbersome and tedious task prone to human error, it is a prerequisite for the training of supervised deep learning approaches that in turn enable accurate automated delineation of fetal brain tissues~\cite{khalili_automatic_2019,hong_fetal_2020,dou_deep_2020,delannoy_segsrgan_2020}.


Despite two recent public releases of annotated datasets \cite{payette_automatic_2021, gholipour_normative_2017} that support the development of accurate automatic deep learning segmentation methods, their generalization across different domains remains barely explored~\cite{payette_efficient_2020}. Specifically, the training of a deep learning network relies on an optimization problem to fit a data distribution, and tends to perform poorly on a new dataset with a different distribution due to variations of MR vendors, acquisition parameters or image reconstruction techniques. Therefore, while a segmentation framework can apply successfully to a specific dataset, it might fail on a different but related dataset. Domain adaptation 
addresses this data distribution gap by generalizing a model from one domain to another. It consists in using knowledge learned in a source domain to adapt a model to a target domain. Transfer learning has recently proven to be effective in adapting automatic segmentation methods for fetal brain MRI from different reconstruction methods~\cite{payette_efficient_2020}. 

In this paper, we explore different domains due to different SR reconstruction methods, as it often occurs when dealing with datasets from multiple hospitals and research teams. We thus study how synthetic MR images, which mimic our data target domain, can help to adapt a tissue segmentation network trained from another domain.


\begin{figure*}[h!]
    \centering
    \includegraphics[width=0.9\linewidth]{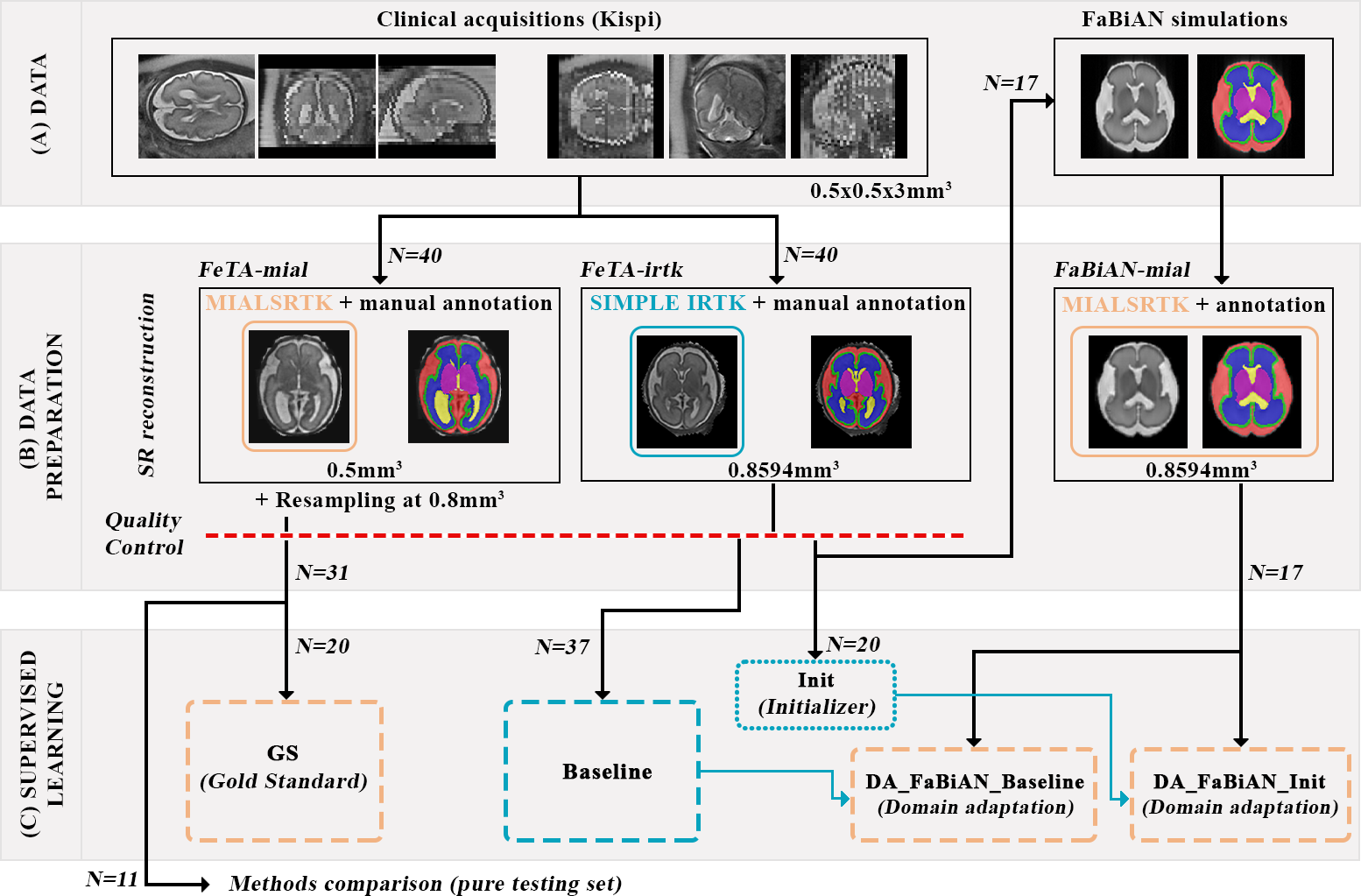}
    \caption{Overall framework. Panels (A) and (B) illustrate the clinical and simulated acquisitions and the preprocessing steps. Panel (C) summarizes the different configurations that are evaluated.}
\label{fig:experiment_design}
\end{figure*}

\section{Methodology}
The framework displayed in Fig.~\ref{fig:experiment_design} provides an overview of the datasets (A and B) and the experimental design (C).

\subsection{Domain gap}
\begin{figure}[h!]
    \centering
    \includegraphics[width=0.9\linewidth]{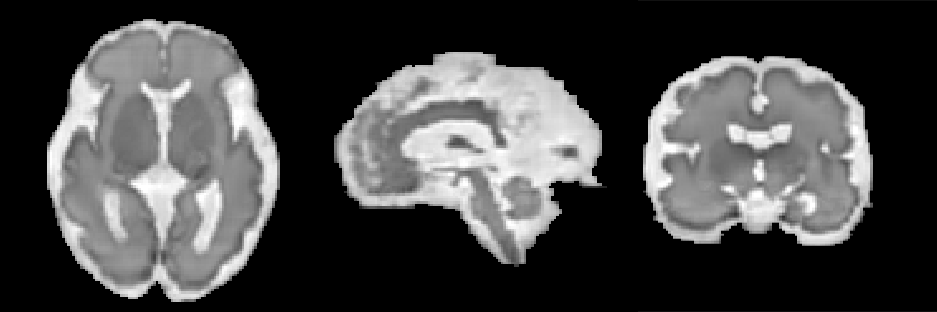}
    \caption{Illustration of \textit{FaBiAN-mial} target-like domain.}
\label{fig:Fab_domain}
\end{figure}

SR reconstruction techniques rely on several pre-processing steps (denoising, bias field correction, intensity normalization, etc.), and different direct model formulation and regularization approaches. This leads to SR reconstructed images that might be very different in terms of intensity and sharpness, as previously discussed in~\cite{payette_efficient_2020}, and thus generate different domains.
We define the following domains (see Fig.~\ref{fig:Fab_domain}):
\begin{description}
\item[Source domain] SR obtained with a simplified version of the Image Registration Toolkit (SIMPLE IRTK)~\cite{kuklisova-murgasova_reconstruction_2012} under Licence from Ixico Ltd (\textit{FeTA-irtk}), 
\item[Target domain] MIALSRTK pipeline~\cite{tourbier_efficient_2015,tourbier_sebastientourbiermialsuperresolutiontoolkit_2019} (\textit{FeTA-mial}),
\item[Target-like domain] Numerically-simulated images SR-reconstructed with MIALSRTK (\textit{FaBiAN-mial}).
\end{description}

\subsection{Clinical datasets (\textit{FeTA-irtk} and \textit{FeTA-mial})}
Among the 80 subjects in the publicly available Fetal Tissue Annotation Dataset (FeTA 2.0)~\cite{payette_automatic_2021}, 40 were reconstructed with an isotropic resolution of $0.8594 mm$ (\textit{FeTA-irtk}) and 40 with an isotropic resolution of $0.5 mm$ (\textit{FeTA-mial}). The SR images of this latter set was resampled to an isotropic resolution of $0.8 mm$ and annotations were refined~\cite{fidon_label-set_2021}.
The extra-axial cerebrospinal fluid spaces (CSF), the cortical gray matter (GM), the white matter (WM), the ventricular system (lateral, third and fourth ventricles), the cerebellum, the deep gray matter (dGM) and the brain stem were manually annotated in the resulting \textit{FeTA-irtk} and \textit{FeTA-mial} SR volumes.
An engineer with 20 years of experience in medical image processing assessed the quality of the 3D SR reconstructions. Only subjects with excellent (good quality without any blurring) and acceptable (overall good quality with some blurring but still relevant for diagnosis purposes) SR were considered in the following.

Thirty-one subjects (15 neurotypical and 16 pathological subjects in the gestational age (GA) range of 20.0 to 33.4 weeks) from \textit{FeTA-mial} and 37 subjects (17 neurotypical and 20 pathological subjects in the GA range of 20.1 to 34.8 weeks) from \textit{FeTA-irtk} were kept after quality control.



\subsection{Simulated data in target-like domain (\textit{FaBiAN-mial})}
The SR label maps from 17 clinical cases (nine neurotypical and eight pathological subjects in the GA range of 20.9 to 34.8 weeks) among the 37 \textit{FeTA-irtk} subjects serve as fetal brain models to generate realistic synthetic images of the fetal brain throughout development.
Typical T2-weighted single-shot fast spin echo (SS-FSE) acquisitions of the fetal brain are simulated at either 1.5 T or 3 T using FaBiAN, a Fetal Brain magnetic resonance Acquisition Numerical phantom~\cite{lajous_fabian_2021,lajous_fabian_zenodo_2021}.
For every subject, partially-overlapping series of 2D thick slices are simulated in each of the three orthogonal orientations with random little or moderate motion.

The simulated LR images are interpolated to $0.8594 \times 0.8594 mm^2$ in the in-plane direction to match the resolution of the clinical SR reconstructions. Automated segmentation of the fetal brain is propagated in all LR image series from the initial SR label maps by a nearest-neighbour interpolation using FaBiAN simulation framework~\cite{lajous_fabian_zenodo_2021}. The fetal brain and its corresponding segmentation are SR-reconstructed using the docker version of the MIALSRTK pipeline~\cite{tourbier_efficient_2015,tourbier_mialsuperresolutiontoolkit_2020}, adapted to reconstruct associated label maps (\textit{FaBiAN-mial}).

\begin{figure*}[ht!]
    \centering
    \includegraphics[width=\linewidth]{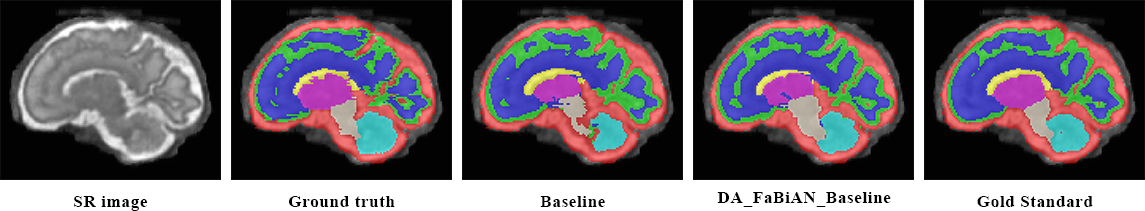}
    \caption{Sagittal view of the segmentation obtained in the different configurations for a fetus of 31.2 weeks of GA.}
\label{fig:segmentation_results}
\end{figure*}

\subsection{Segmentation networks}

\textbf{Configurations.}
The popular 2D U-Net~\cite{Ronneberger_Unet_2015} architecture is selected as it recently proved its ability to deal with fetal brain MRI tissue segmentation~\cite{khalili_automatic_2019, payette_efficient_2020}. It is run in the different training configurations described below (Fig.~\ref{fig:experiment_design}-panel C):

\begin{itemize}
    \item A \textbf{Gold Gtantard} (GS) network is trained using 20 randomly selected subjects from the \textit{FeTA-mial} target domain (N=20, target domain).
    
    \item A \textbf{Baseline} network is trained on the full \textit{FeTA-irtk} set (N=37, source domain).  
    
    \item An \textbf{Initializer} (Init) network is trained on the 20 \textit{FeTA-irtk} subjects that are not used for simulation purposes (N=20, source domain). 
    
    \item \textbf{DA\_FaBiAN\_Baseline} and \textbf{DA\_FaBiAN\_Init} are initialized respectively with Baseline and Init pre-trained weights to perform transfer learning using the 17 \textit{FaBiAN-mial} subjects (N=17, target-like domain).
\end{itemize}

A pure testing set that consists of the 11 remaining \textit{FeTA-mial} subjects (five neurotypical and six pathological subjects) is used to compare the models.


\textbf{Strategy.}
The networks are trained on $64 \times 64$ voxel size image patches using a multi-view approach, considering intracranial voxels only. All patches are duplicated and the following data augmentation is randomly applied: rotation,
flipping and addition of Gaussian noise. Intensities of all patches are standardized. 
A cross-validation strategy is adopted to tune the learning rate decay. All networks are trained by optimizing a hybrid loss function that combines, with equal contribution, a categorical cross-entropy loss and a Dice similarity coefficient (DSC).

At test-time, overlapping patches are inferred from the three orientations and final prediction is reconstructed using a majority voting strategy.

\textbf{Evaluation.}
The performance of the networks is evaluated with the DSC \cite{dice_1945} between the ground truth manual annotations and the predicted segmentation. A paired Wilcoxon rank-sum test is performed between each experimental configuration (DA\_FaBiAN\_Init and DA\_FaBiAN\_Baseline) and the Baseline. For individual fetal brain tissues, \textit{p}-values are adjusted for multiple comparisons using Bonferroni correction. Statistical significance level is set to 0.05.

\section{Results}

\begin{table*}[hb!]
  \centering
\begin{tabular}{r || c c c | c }
Network & Baseline & DA\_FaBiAN\_Init & DA\_FaBiAN\_Baseline & Gold Standard \\
\hline
\hline
CSF                 & 0.75 $\pm$ 0.32   & 0.76 $\pm$ 0.33       & \textbf{0.77 $\pm$ 0.33}      & 0.81 $\pm$ 0.30   \\

Cortical GM         & 0.57 $\pm$ 0.22   & 0.63 $\pm$ 0.21       & \textbf{0.64 $\pm$ 0.21} (*)  & 0.68 $\pm$ 0.18   \\

WM                  & 0.77 $\pm$ 0.22   & 0.79 $\pm$ 0.21       & \textbf{0.80 $\pm$ 0.21} (*)  & 0.86 $\pm$ 0.19   \\

Ventricles          & \textbf{0.79 $\pm$ 0.20}   & 0.75 $\pm$ 0.20       & 0.76 $\pm$ 0.23      & 0.85 $\pm$ 0.17   \\

Cerebellum          & 0.58 $\pm$ 0.30   & \textbf{0.71 $\pm$ 0.35} (*)  & 0.70 $\pm$ 0.35 (*)   & 0.65 $\pm$ 0.32   \\

Deep GM             & 0.48 $\pm$ 0.20   & 0.57 $\pm$ 0.21       &  \textbf{0.59 $\pm$ 0.22} (*) & 0.82 $\pm$ 0.19   \\

Brain stem          & 0.53 $\pm$ 0.28   &  \textbf{0.67 $\pm$ 0.23} (*)  & 0.65 $\pm$ 0.25 (*)  & 0.76 $\pm$ 0.26   \\

\hline

Overall             & 0.64 $\pm$ 0.23   & 0.70 $\pm$ 0.2 (*)    & \textbf{0.70 $\pm$ 0.24} (*)  & 0.77 $\pm$ 0.22   \\

\end{tabular}

\caption{\label{tab:DSC_tissue_overall}
DSC (mean $\pm$ standard deviation) of different training configurations. The best scores between domain adaptation (DA\_FaBiAN) configurations and the Baseline are shown in bold. The corresponding \emph{p}-values (paired Wilcoxon rank sum test) are adjusted for multiple comparisons using Bonferroni correction. \emph{p} $< 0.05$ (*) is considered statistically significant.
}
\end{table*}

\begin{figure*}[hb!]
    \centering
    \includegraphics[width=0.95\linewidth]{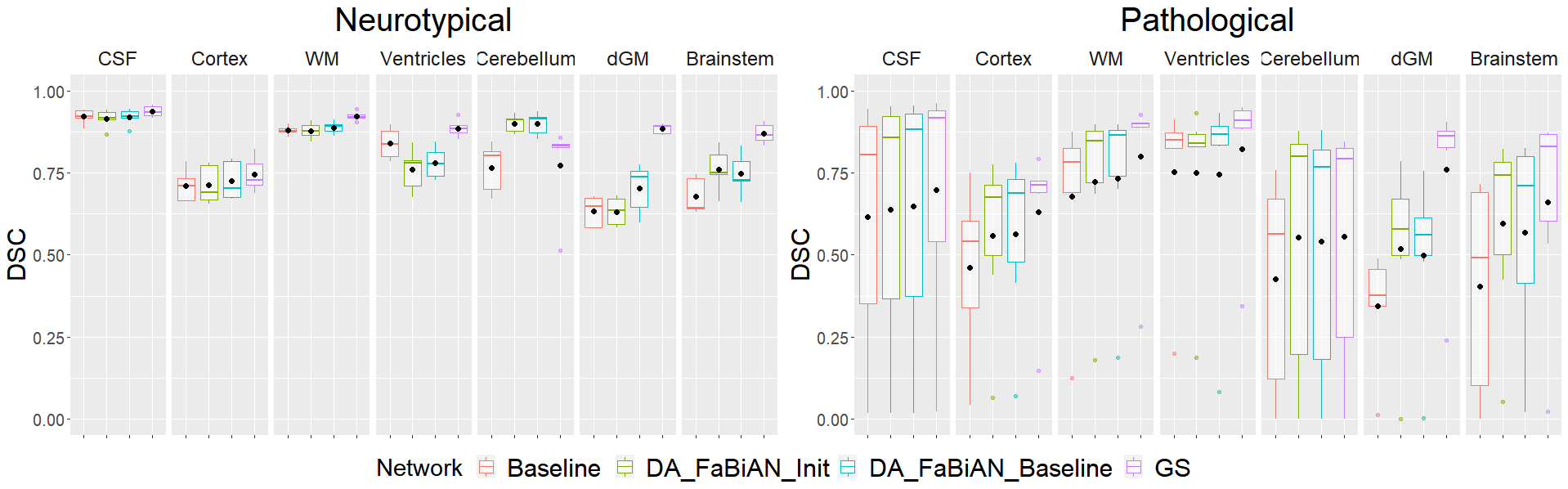}
    \caption{Label-wise inter-method comparison for neurotypical and pathological subjects.}
\label{fig:boxplot_NT_PT}
\end{figure*}

Qualitative results are shown in Fig.~\ref{fig:segmentation_results} for a fetus of 31.2 weeks of GA. Table \ref{tab:DSC_tissue_overall} shows the mean DSC $\pm$ standard deviation computed for every segmented fetal brain tissue in each configuration studied. Overall, the performance of the segmentation algorithm is significantly enhanced when fine-tuning the Baseline network with synthetic, yet realistic fetal brain MR images reconstructed in the target-like domain (DA\_FaBiAN). 
The accuracy of the segmentation algorithm is slightly decreased in the ventricles only, but without statistical significance. This may be due to erroneous label propagation (after simulation and reconstruction) resulting in neighboring areas of similar intensities at the mid-sagittal plane either segmented as CSF or ventricles. Conversely, the segmentation of the cerebellum is even more accurate when complementing the Baseline dataset with synthetic images (DA\_FaBiAN) than when training the network directly with the target \textit{FeTA-mial} dataset (Gold Standard).

Interestingly, our domain adaptation strategy leads to a more accurate segmentation of the CSF, the cortical GM, the WM, the cerebellum, the deep GM and the brain stem in pathological cases than the Baseline, as shown in Fig.~\ref{fig:boxplot_NT_PT}. It is also the case for the ventricles when combining domain adaptation and data augmentation approaches \linebreak (DA\_FaBiAN\_Baseline). The trend is different in neurotypical subjects, where DA\_FaBiAN\_Baseline only results in an improved segmentation of the deep GM, the brain stem and the cerebellum. The segmentation of the latter is even more accurate than when training the network on clinical data in the target domain (GS). However, the performance of the segmentation algorithm is stable for the CSF and the cortex, with a slight increase for the WM in the configuration DA\_FaBiAN\_Baseline, whereas it decreases for the ventricles.

\section{Conclusion}
We have demonstrated for the first time that synthetic \linebreak numerically-generated MR images of the developing fetal brain can be used for transfer learning between two different domains. The proposed domain adaptation strategy has proven to be powerful since it is free of human intervention in the sense that it does not require expert annotations in the target domain. Thus, it overcomes the lack of clinical annotations and can be generated from any pre-existing label map. Overall, our method significantly enhances the performance of the baseline fetal brain tissue segmentation algorithm.
Future work will further extend the idea of domain adaptation to multi-scanner multi-center datasets.


\clearpage

\section{Compliance with ethical standards}
\label{sec:ethics}

This research study was conducted retrospectively using human subject data from the open access Fetal Tissue Annotation and Segmentation Dataset (FeTA).
These subjects were scanned in accordance with the relevant guidelines and regulations and their inclusion in research studies was approved by the ethical committee of the Canton of Zurich, Switzerland (KEK, decision number: 2016-01019).\\

\section{Acknowledgments}
\label{sec:acknowledgments}

This work is supported by the Swiss National Science Foundation through grants 182602 and 141283. We acknowledge access to the facilities and expertise of the CIBM Center for Biomedical Imaging, a Swiss research center of excellence founded and supported by Lausanne University Hospital (CHUV), University of Lausanne (UNIL), Ecole polytechnique fédérale de Lausanne (EPFL), University of Geneva (UNIGE) and Geneva University Hospitals (HUG).

The authors have no relevant financial or non-financial interests to disclose.

\bibliographystyle{IEEEbib}
\bibliography{refs}

\end{document}